\documentclass[ruled,linesnumbered,noline,
final
]{dmtcs-episciences}

\usepackage{amsthm}
\usepackage{enumerate}

\usepackage{amsmath,capt-of}
\usepackage{listings}
\usepackage{tikz}
\usetikzlibrary{shapes.geometric, arrows,oodgraph}
\usetikzlibrary{shapes.arrows}
\usetikzlibrary{decorations.pathreplacing}

\newcommand{\ignore}[1]{}

\DeclareMathOperator{\fact}{factors}
\newcommand{\factors}[1]{\fact\left(#1\right)}
\newcommand{\structure}[1]{\ensuremath{\left(#1\right)}\xspace}
\newcommand{\isomorfo}{\approx}
\newcommand{\set}[1]{\ensuremath{\left\{#1\right\}}\xspace}

\usepackage{complexity}
\newclass{\EnumP}{EnumP\xspace}
\newlang{\CMP}{CMP\xspace }
\newlang{\SOBFID}{SOBFID\xspace}
\newlang{\ESOBFID}{EnumSOBFID\xspace}
\newlang{\DEEP}{DEEP\xspace}

\usepackage[ruled,vlined]{algorithm2e}

\newcommand{\ie}{\emph{i.e.,}\@\xspace}

\usepackage{algpseudocode}

\def\newv#1{#1}

\usepackage{mathtools}

\definecolor{bleu}{rgb}{0, 0.6, 0.8}
\definecolor{rose}{rgb}{0.8, 0, 0.4}
\definecolor{vert}{rgb}{0, 0.6, 0.4}

\tikzstyle{condition}  = [diamond, draw, fill=bleu!50, text width=5em, text badly centered, inner sep=0pt]
\tikzstyle{conditionFunction}  = [diamond, draw, fill=white, text width=5em, text badly centered, inner sep=0pt]
\tikzstyle{solver}     = [draw, ellipse,fill=white, text centered,  minimum height=2em,text width=3.7em]
\tikzstyle{block}      = [rectangle, draw, fill=bleu!80, text width=7em, text centered, rounded corners, minimum height=2em]
\tikzstyle{bigblock}   = [rectangle, draw, fill=bleu!80, text width=8em, text centered, rounded corners, minimum height=2em]
\tikzstyle{line}       = [draw, -latex']
\tikzstyle{problem}    = [circle,double, draw, fill=black!50, text width=3em, text centered, rounded corners, minimum height=1em]
\tikzstyle{emptyblock} = [rectangle, fill=white, text centered, text width=2em, rounded corners, minimum height=2em]
\tikzstyle{emptybigblock} = [rectangle, fill=white, text width=4em, rounded corners, minimum height=2em]
\tikzstyle{vertex}=[circle, draw]
\usetikzlibrary{shapes.multipart}
\usetikzlibrary{shapes,arrows,oodgraph}

\newtheorem{definition}{Definition}
\newtheorem{lemma}{Lemma}
\newtheorem{theorem}{Theorem}
\newtheorem{proposition}{Proposition}

\newtheorem{example}{Example}
\newtheorem*{problem}{Problem}
\newcommand{\N}{\mathbb{N}\xspace}
\DeclareMathOperator{\lcm}{lcm}
\usetikzlibrary{decorations.markings}
\usepackage{tabularx}

\newcommand{\ring}{R}
\newcommand{\lcmm}{\lambda}
\newcommand{\nP}{l}
\newcommand{\basep}{\mathsf{p}}
\newcommand{\espp}{h}
\usepackage[round]{natbib}

\author[Alberto Dennunzio et al.]{
Alberto Dennunzio\affiliationmark{1}
  \and Enrico Formenti\affiliationmark{2}
  \and Luciano Margara\affiliationmark{3}
  \and Sara Riva\affiliationmark{4}
  }

\title[On solving basic equations over the semiring of functional digraphs]{On solving basic equations over the semiring of functional digraphs}
\affiliation{
Dipartimento di Informatica, Sistemistica e Comunicazione,
  Università degli Studi di Milano-Bicocca,
  Milano, Italy\\
Université Côte d'Azur, CNRS, I3S, Nice, France\\
Department of Computer Science and Engineering, University of Bologna, Cesena Campus, Cesena, Italy\\
Univ. Lille, CNRS, Centrale Lille, UMR 9189 CRIStAL, Lille, France
}
\begin{document}
\publicationdata{vol. 27:3}{2025}{6}{10.46298/dmtcs.14590}{2024-10-17; 2024-10-17; 2025-05-26; 2025-07-16}{2025-07-23}
\maketitle         
\begin{abstract}\mbox{}\\
Endowing the set of functional digraphs with the
sum (disjoint union of digraphs) and product (standard direct product on digraphs) operations
induces on FDs a structure
of a commutative semiring $\ring$.
The operations on $\ring$ can be naturally extended to the set of univariate polynomials $\ring[X]$
over $\ring$. This paper provides
a polynomial time algorithm
for deciding if equations of the type $AX=B$ have solutions
when $A$ is just a single cycle
and $B$ a set of cycles of identical
size. We also prove a similar complexity result for some variants of the previous equation.
\end{abstract}
\section{Introduction}
A \textbf{functional digraph} (FD) is the digraph of a function with finite domain \ie a digraph with outgoing degree $1$. Endowing the set of functional digraphs FGs with the
sum (disjoint union of digraphs) and product (standard direct product on digraphs) operations
provides FGs with the structure
of a commutative semiring $\ring$ in which
the empty digraph (resp., the single loop) is the neutral element of addition (resp., of
product)~\cite{dorigatti2018}. This semiring
can be naturally extended to the semiring
of multivariate polynomials $\ring[X_1,X_2,\ldots, X_k]$.

A polynomial in $\ring[X_1,X_2,\ldots, X_k]$
represents a (infinite) set of functions
which have a common substructure provided
by the coefficients (modulo isomorphism). The factorization is an
interesting inverse problem in this context.
Indeed, assume to have a functional digraph
$G$. Can $G$ be factorized \ie decomposed
into the product or the sum (or a combination
of the two) of functional digraphs with smaller vertex set\footnote{We stress that
factorization here is meant in the context of abstract algebra which has nothing to
do with the notion of factor graph often used in graph theory.}? If this question has
been largely investigated for general graphs (see, for instance, \cite{weichsel1962,abay-asmeron2010,hammack2011handbook}),
little is known for functional digraphs.

Let us consider another version of the
factorization problem. Assume that we have
partial information (or partial assumptions)
on the factorization of $G$. This partial
information is represented by a sequence
of functional digraphs $A_1, A_2, \ldots, A_k$ and the question of factorization
can be reformulated as ``does the following
equation 
\[
A_1\cdot X_1+A_2\cdot X_2+\ldots+A_k\cdot  X_k
= G
\]
admit a solution?'' \ And more generally, one
can ask for solutions of the following
\begin{equation}
\label{eq:poly=costante}
A_1\cdot X_1^{w_1}+A_2\cdot X_2^{w_2}+\ldots+A_k\cdot X_k^{w_k}
= G
\end{equation}

where $X^w$, as usual, is the multiplication
of $X$ with itself $w$ times. Figure~\ref{fig:generic-poly-equation} provides an example of equation between two multivariate polynomials and one of its solutions.

In~\cite{dorigatti2018}, it is shown that
the problem of deciding if there exist solutions to
Equation~\eqref{eq:poly=costante} is in
\NP. However, we do not know if the problem is \NP-complete.
Our current conjecture is that it might be in \P.
In~\cite{DENNUNZIO2023101932}, an algorithmic pipeline is provided for finding
the solutions of~\eqref{eq:poly=costante}
when all the digraphs involved are the digraph
of a permutation (\ie they are unions of
loops and cycles).
The software pipeline essentially
relies on the solution of a finite (potentially
exponential) number of \textbf{basic equations} of the type
\begin{equation}\label{eq:basic}
AX=B
\end{equation}
where $A$ is a single cycle of size $p$
and $B$ is a union of cycles of size $q$ (of course, it might be $p\ne q$). In this paper, we prove
that deciding if equations of
type~\eqref{eq:basic} have solutions or not is in \P.

The interest in this result is manyfold. Without a doubt, it will
significantly improve the performance of the software
pipeline in ~\cite{DENNUNZIO2023101932} for the case in which one is interested in the existence of solutions and not in their enumeration.
%\sr{Non ci sarebbe miglioria diretta... bisogna specificare che l'algoritmo proposto enumera e che quindi si potrebbe proporne una nuova diversa versione se si è interessati all'esistenza della soluzione} 
It will shed some
light on some connected important questions concerning the cancellation problem for functional digraphs \ie the
problem of establishing if $G\cdot F\isomorfo G\cdot H$ implies $F\isomorfo H$ where $F, G$ and
$H$ are FDs and $\isomorfo$ is the relation of graph isomorphism~\cite{DORE2024114514, NAQUIN2024114509}. We refer the reader to
the classical book of \cite{hammack2011handbook} for the factorization and cancelation problems for more general graphs.

%
% !TEX root = main.tex
%%%%%%%%%%%%%%%%%%%%%%%%%%%%%%%%%%%%%%%%%%%%%%%%%%%%%%%%%%%%%%%%%%%%%%%%%%%%%%%%%%%%%%%%%%%%%%%%%%%%%%%%%%%%%%%%%%%%%%%%%%%%%%%%%%%%%%%%%%%%%%%%%%

\section{Background and basic facts}\label{back}
%
%\mytodo{Comincio da qui}

\ignore{
A \textbf{Finite Discrete Dynamical System (FDDS)} is a pair $\structure{\setst,f}$ where $\setst$ is a finite set of \textbf{states} and 
$f: \setst \to \setst$ is the \textbf{next state map}
which associated the current state of the system with
the new one after one time step. Any FDDS $\structure{\setst,f}$ can be identified with its 
\textbf{dynamics graph} $G=\structure{V,E}$, where $V=\setst$ and $E=\set{(\nodedds,f(\nodedds))\, |\, \nodedds\in V}$ is the graph of $f$.
%\smallskip

Let $\dds$ be a FDDS $\structure{\setst,f}$ and let $G$ be its dynamics graph.  If $\mathcal{Y}$ is any subset of $\setst$ such that $f(\mathcal{Y})\subseteq \mathcal{Y}$, then 
the FDDS $\structure{\mathcal{Y},f|_\mathcal{Y}}$ is the \textbf{sub-dynamical system} of $\structure{\setst,f}$ induced by $\mathcal{Y}$ (here, $f|_\mathcal{Y}$ means the restriction of $f$ to $\mathcal{Y}$). Clearly, the dynamics graph of $\structure{\mathcal{Y},f|_\mathcal{Y}}$ is nothing but the subgraph of $G$ induced by $\mathcal{Y}$. 
A state $\nodedds\in\setst$ is \textbf{periodic} for $\dds$ if there exists an integer $ p>0$ such that $f^{p}(\nodedds)=\nodedds$. The smallest $p$ with the previous property is the \textbf{period} of $\nodedds$. If $p=1$, the state $\nodedds$ is a \textbf{fixed point}. A \textbf{cycle} (of length $p$) of $\dds$ is a set $\mathcal{C}=\{\nodedds,f(\nodedds), ..., f^{p-1}(\nodedds)\}$ where $\nodedds\in\setst$ is a periodic point of period $p$. Clearly, the set $\pp$ of all the periodic points of $\dds$ can be viewed as an union of disjoint cycles. Moreover, the
FDDS $\structure{\mathcal{C},f|_\mathcal{C}}$ (resp., $\structure{\pp,f|_\pp}$) is a sub-dynamical system of $\dds$ and its dynamics graphs consists of one of
(resp., all) the strongly connected components of $G$.  
In the sequel, we will identify $\mathcal{C}$ and $\pp$ with the FDDS $\structure{\mathcal{C},f|_\mathcal{C}}$ and $\structure{\pp,f|_\pp}$ (and then with their dynamics graphs too), respectively.

\smallskip
Two FDDS are
\textbf{isomorphic} 
if their dynamics graph are isomorphic in the usual sense of graph theory. 
When this happens, the two systems are indistinguishable from the dynamical systems point of view. In particular, periodic points and cycles of the two systems are in one-to-one correspondence. Therefore, the dynamical sub-systems induced by them in their respective FDDS are isomorphic too.

\smallskip
%From now on, we will identify a DDS with its dynamics graph.
}

In~\cite{dorigatti2018}, an abstract algebraic setting for
studying finite discrete dynamical systems was introduced. This setting, can be adapted in a straightforward way to FDs. In this section, we recall just the minimal concepts to understand the problem we want to solve.

\begin{definition}[Sum of FDs]\label{ddssum}
    For any pair of FGs $F=\structure{V_F, E_F}$
    and
    $G=\structure{V_G, E_G}$, the \textbf{sum} $H=\structure{V_H, E_H}$ of $F$ and $G$
    is defined as
\begin{align*}
V_H&=V_F\sqcup V_G\\
E_H&=\set{((a,0),(b,0)) \text{ s.t. } (a,b)\in E_F}\cup\set{((a,1),(b,1))
\text{ s.t. } (a,b)\in E_G}   
\end{align*}
where $A\sqcup B$ is the disjoint
union of $A$ and $B$ defined as
$A\sqcup B=\left(A\times\set{0}\right)\cup\left(B\times\set{1}\right)$.
\end{definition}
The product of FDs is the standard direct product of graphs that we recall here for completeness sake.
\begin{definition}[Product of FDs]
For any pair of FGs $F=\structure{V_F, E_F}$ and
$G=\structure{V_G, E_G}$, the \textbf{product} $H=\structure{V_H, E_H}$ of $F$ and $G$ is defined as
\begin{align*}
    V_H&=V_F\times V_G\\
    E_H&=\set{\left((a,c),(b,d)\right)\in V_H\times V_H\text{ s.t }(a,b)\in E_F\text{ and }(c,d)\in E_G}
\end{align*}
\end{definition}
%Recall that the \textbf{disjoint union} of two sets $\setst_1$ and $\setst_2$ is the set $\setst_1 \sqcup \setst_2 = (\setst_1 \times \{0\}) \cup (\setst_2 \times \{1\})$. This notion can be extended to functions in
%the obvious way.
When no misunderstanding is possible, we will denote with $+$ (resp., $\cdot$) the sum (resp., the product) of FDs.
\smallskip

As already informally stated in
the introduction, we have the following algebraic characterisation of the class of FDs.

\begin{theorem}[\cite{dorigatti2018}]
   The class of FDs (modulo isomorphisms) equipped with
   the operations of $+$ and $\cdot$
   is a commutative semiring.
\end{theorem}
\begin{figure}
\begin{center}
\begin{tikzpicture}
\begin{oodgraph}[scale=.5]
\addcycle{3};
\end{oodgraph}
\node[font=\Large,anchor=west] at (.55,.08) {$\cdot\; X+Y^2=$};
\begin{oodgraph}[xshift=3.25cm,yshift=.3cm, rotate=-45, scale=.75]
\addcycle[nodes prefix=mybeard]{2};
\addbeard[attach node=mybeard->1]{1};
\end{oodgraph}
\node[font=\Large,anchor=west] at (3.55,.075) {$\cdot\;Z+$};
\begin{oodgraph}[xshift=5.25cm,rotate=90]
\addcycle[nodes prefix=mybeard,rotate=-90]{1};
\addbeard[attach node=mybeard->1]{3};
\end{oodgraph}
%\end{tikzpicture}
%\begin{tikzpicture}
\begin{scope}[yshift=-2cm]
\node[anchor=west,font=\Large] at (0,0) {$X=$};
\begin{oodgraph}[xshift=1.45cm,yshift=.25cm, rotate=-45, scale=.75]
\addcycle[nodes prefix=mybeard]{2};
\addbeard[attach node=mybeard->1]{1};
\end{oodgraph}
\node[anchor=west,font=\Large] at (2.5,0) {$Y=$};
\begin{oodgraph}[xshift=3.95cm,yshift=.04cm, rotate=-90, scale=.75]
\addcycle[nodes prefix=mybeard,rotate=-90]{1};
\addbeard[attach node=mybeard->1]{1};
\end{oodgraph}
\node[anchor=west,font=\Large] at (4.9,0) {$Z=$};
\begin{oodgraph}[xshift=6.5cm,yshift=.125cm, scale=.5]
\addcycle[nodes prefix=mybeard,rotate=-90]{3};
\end{oodgraph}
\end{scope}
\end{tikzpicture}
\end{center}
\caption{An example of polynomial equation over FDs (above) and a solution (below).}
\label{fig:generic-poly-equation}
\end{figure}

We also
recall the following important result on the complexity of
solving polynomial equations.
\begin{theorem}[\cite{dorigatti2018}]
The problem of deciding if there are
solutions to
\begin{itemize}
    \item polynomial equations over $\ring[X_1, X_2, \ldots]$ is undecidable;
    \item polynomial equations over $\ring[X_1, X_2, \ldots]$ with constant right-hand side is in \NP.
\end{itemize}
\end{theorem}
We remark that the problem of finding solutions for polynomial equations with a constant right-hand term
has some similarities with other well-known \NP-complete problems
(knapsack, change-making, for instance). However, we do not know
if it is complete or not.
This further motivates the quest for simpler equations which admit solution algorithms with lower complexity.
These algorithms will be very convenient for use in practical applications. Indeed, in biology, when analyzing the behavior of a gene regulation network,for example, one can ask if the observed behavior
is produced by the network at hand or if it is
the result of the cooperation of simpler systems.
These questions can be easily translated in terms
of equations over FDs. However, gene regulation
networks may contain hundreds of genes and hence
effective methods are necessary to solve those
equations.

\section{Solving basic equations on permutations}\label{sec:a-abstraction}

From now on, we will focus on a subclass of functional digraphs, namely, the digraphs of permutations.
For this reason, we introduce a convenient notation called \textbf{C-notation} inspired from~\cite{DENNUNZIO2023101932}.
We note $C_p$ the graph made
by a single cycle of size $p$.
Similarly, $n\cdot C_p$ (or simply $nC_p$) denotes a graph
that is the sum of $n$ graphs
$C_p$. According to this notation, it is clear that $aC_p+ bC_p =(a+b)C_{p}$ and $a\cdot bC_p=abC_p$ for any natural $a,b$. With this notation, the classical result that the graph $G$ of a permutation is a union of disjoint
cycles translates into
\[
G=\sum_{i=1}^l n_iC_{p_i}
\]
for suitable positive integers $p_1,\ldots,p_l$ and $n_1,\ldots,n_l$. Figure~\ref{fig:exemple-operations-c-notation} provides an example of operations on permutation digraphs and their expression through the C-notation.
\begin{figure}
    \begin{center}
        \begin{tikzpicture}[scale=.5]
            \begin{oodgraph}[/oodgraph/nodes color=red]
                \addcycle{3};
            \end{oodgraph}
            \begin{oodgraph}[/oodgraph/nodes color=blue,xshift=3.5cm]
                \addcycle{2};
            \end{oodgraph}
            \node[font=\Large] at (2,0) {$+$};
            \node[font=\Huge] at (4.5,0) {$)$};
            \node[font=\Huge] at (-1.25,0) {$($};
            \node[font=\Large] at (-2.25,0) {$\cdot$};
            \begin{oodgraph}[/oodgraph/nodes color=green,xshift=-3.75cm]
                \addcycle{2};
            \end{oodgraph}
            \node[font=\Large] at (5.65,0) {$=$};
            \begin{oodgraph}[xshift=7.75cm]
            \addcycle{6};                
            \end{oodgraph}
            \node[font=\Large] at (9.75,0) {$+$};
            \begin{oodgraph}[xshift=11.5cm,yshift=.75cm]
                \addcycle{2};
            \end{oodgraph}
            \begin{oodgraph}[xshift=11.5cm,yshift=-.75cm]
                \addcycle{2};
            \end{oodgraph}
        \end{tikzpicture}
    \end{center}
    \caption{Example of operations on FDs of permutations. Using the C-notation, we have $\color{green}C_2\color{black}\cdot(\color{red}C_3\color{black}+\color{blue}C_2)\color{black}=\color{green}C_2\color{black}\cdot\color{red}C_3\color{black}+\color{green}C_2\color{black}\cdot\color{blue}C_2\color{black}=C_6+2C_2$.}
    \label{fig:exemple-operations-c-notation}
\end{figure}

\ignore{
\definecolor{verde}{rgb}{0,0.4,0.4}
\begin{figure}[]
	\centering
	\includegraphics[scale=1.2, keepaspectratio]{exAabs.pdf}
	\caption{A DDS with four cycles ($\nP=3$): $(\textcolor{blue}{C^1_1} \oplus \textcolor{red}{C^2_2} \oplus \textcolor{verde}{C^1_3})$ in our notation. }
	\label{exass}
\end{figure}
}
The following proposition provides an explicit expression for the product of unions of cycles.
\begin{proposition}\label{prop:prod}
For any natural $l>1$ and any positive naturals $n_1,\ldots, n_l$, $p_1,\ldots, p_l$, it holds
\begin{equation*}
\begin{aligned}
\prod\limits_{i=1}^ln_iC_{p_i} = \frac{\tilde{p}_l\tilde{n}_l}{\lcmm_l} C_{\lcmm_l}\enspace.
\end{aligned}
 \end{equation*}
where $\lcmm_l=\lcm(p_1, \ldots, p_l)$, $\tilde{p}_l=\prod_{i=1}^lp_i$ and $\tilde{n}_l=\prod_{i=1}^ln_i$.
\end{proposition}
\begin{proof}
We proceed by induction over $l$.  
First of all, we prove that the statement is true for $l=2$, \ie
\begin{equation}
\label{due}
n_1C_{p_1}\cdot n_2C_{p_2} = \frac{(p_1p_2)(n_1n_2)}{\lcmm_2}C_{\lcmm_2}\enspace.
\end{equation}
Consider the case $n_1=n_2=1$. Since $C_{p_1}$ and $C_{p_2}$ can be viewed as finite cyclic groups of order $p_1$ and $p_2$, respectively, each element of the product of such cyclic groups has order $\lcmm_2=\lcm(p_1, p_2)$ or, in other words, each element of $C_{p_1}\cdot C_{p_2}$ belongs to some cycle of length $\lcmm_2$. So, $C_{p_1}\cdot C_{p_2}$ consists only of $(p_1\cdot p_2)/\lcmm_2$ cycles, all of length $\lcmm_2$, and therefore
\[
C_{p_1}\cdot C_{p_2}=\frac{p_1p_2}{\lcmm_2}C_{\lcmm_2}\enspace.
\]
In the case $n_1\neq 1$ or $n_2\neq 1$, since the product is distributive over the sum, we get
\[
n_1C_{p_1}\cdot n_2C_{p_2} = \sum\limits_{i=1}^{n_1} C_{p_1} \cdot\; \sum\limits_{j=1}^{n_2}  C_{p_2} = \sum\limits_{i=1}^{n_1} \sum\limits_{j=1}^{n_2}  (C_{p_1} \cdot\; C_{p_2})=
\sum\limits_{i=1}^{n_1} \sum\limits_{j=1}^{n_2}  \frac{p_1p_2}{\lcmm_2}C_{\lcmm_2}=
\frac{(p_1p_2)(n_1n_2)}{\lcmm_2}C_{\lcmm_2}\enspace.
\]
Assume now that the equality holds for any $\nP>2$. Then, we get
\[
\sum\limits_{i=1}^{l+1}C^{n_i}_{p_i} =
\frac{\tilde{p}_l\tilde{n}_l}{\lcmm_l}C_{\lcmm_l} \cdot\; n_{l+1}C_{p_{l+1}}
= \frac{\tilde{p}_lp_{l+1}\tilde{n}_ln_{l+1}}{\lcm(\lcmm_l, p_{l+1})}C_{\lcm(\lcmm_l, p_{l+1})}
= \frac{\tilde{p}_{l+1}\tilde{n}_{l+1}}{\lcmm_{l+1}}C_{\lcmm_{l+1}}
\enspace,
\]
where $\lcmm_{l+1}=\lcm(\lcmm_l,p_{l+1})$,
$\tilde{p}_{l+1}=\tilde{p}_lp_{l+1}$ and
$\tilde{n}_{l+1}=\tilde{n}_ln_{l+1}$.
\end{proof}

In the sequel, we will also make use of the following notation. Let $n$ be any integer strictly greater than $1$. We note $\factors{n}$ the set of prime
factors of $n$ \ie if $n=n_1^{m_1}\cdot\ldots\cdot n_t^{m_t}$, then $\factors{n}=\set{n_1,\ldots,n_t}$
where $n_1, \ldots, n_t$ are distinct primes.
\medskip

According to the $C$-notation, \textbf{basic equations} will have
the following form
\begin{equation} 
\label{eq:simple}
	C_p\cdot X = nC_q\enspace,
\end{equation}
where $X$ is the unknown and $p,n$ and $q$ are positive integers. We stress that the numbers $p, q$ and $n$ are just positive integers. They should not be thought of as prime integers unless explicitly mentioned. One of the issues of basic equations is that they might admit several distinct solutions. This is because some
digraphs admit several distinct factorizations.
Figure~\ref{fig:multiple-factorizations} provides an easy example of this fact.
\begin{figure}
    \begin{center}
        \begin{tikzpicture}[scale=.5]
            \begin{oodgraph}[/oodgraph/nodes color=orange]
                \addcycle{2};
            \end{oodgraph}
            \node[font=\Large] at (1.5,0) {$\cdot$};
            \begin{oodgraph}[/oodgraph/nodes color=orange,xshift=3cm]
                \addcycle{2};
            \end{oodgraph}
            \node[font=\Large] at (4.5,0) {$=$};
            \node[font=\Large] at (5.25,0) {$($};
            \begin{oodgraph}[/oodgraph/nodes color=blue,xshift=6.5cm]
                \addcycle{1};
            \end{oodgraph}
            \node[font=\Large] at (7.35,0) {$+$};
            \begin{oodgraph}[/oodgraph/nodes color=blue,xshift=8.85cm]
                \addcycle{1};
            \end{oodgraph}
            \node[font=\Large] at (9.35,0) {$)$};
            \node[font=\Large] at (10,0) {$\cdot$};
            \begin{oodgraph}[/oodgraph/nodes color=orange,xshift=11.25cm]
                \addcycle{2};
            \end{oodgraph}
            \node[font=\Large] at (12.7,0) {$=$};
            \begin{oodgraph}[xshift=14.2cm,yshift=.75cm]
                \addcycle{2};
            \end{oodgraph}
            \begin{oodgraph}[xshift=14.2cm,yshift=-.75cm]
                \addcycle{2};
            \end{oodgraph}
        \end{tikzpicture}
    \end{center}
    \caption{Example of functional digraph having multiple factorizations. Using the C-notation, we have $\color{orange}C_2\color{black}\cdot\color{orange}C_2\color{black}=\color{black}(\color{blue}C_1\color{black}+\color{blue}C_1\color{black})\cdot\color{orange}C_2\color{black}=2C_2$.}
    \label{fig:multiple-factorizations}
\end{figure}
\smallskip

As we already stressed in the introduction, in many cases, we are not interested in knowing the exact solutions of a basic equation over permutations but we are interested in deciding if an equation admits a solution or not. %\sr{Qui stesso discorso di prima... io metterei in forma che il problema di enumerazione è già stato studiato ma che anche il problema di decisione è importante (o qualcosa simile).} 
More formally, in this paper, we are interested in
the computational complexity of
the following decision problem.

\begin{problem}[\DEEP, DEciding basic Equations on Permutations]\mbox{}\\
\noindent\hspace*{8pt}\textbf{Instance:} $p,q,n\in\N\setminus\set{0}$.\\
\noindent\hspace*{8pt}\textbf{Question:} Does the equation $C_p\cdot X=nC_q$ admit any solution?
\end{problem}

\DEEP\ is in \NP. In fact, the
number of nodes in the solution $X$ must be a
divisor of $nq$. Hence, one can simply compute
the products with $C_p$ of the disjoint cycles that make up $X$
(using the previous proposition, for example)
and verify that $n$ copies of $C_q$ are obtained.
All this can be clearly done in polynomial time
in the size of $n, p$ and $q$.

However, we are going to show that the problem can be solved
in polynomial time
in Theorem~\ref{th:deep}. 
Remark that this results is stronger than expected. Indeed, as we already said in the introduction, the generic case $AX=B$ is in \NP\ when the input is given in unary (\ie the size of a cycle $C_p$ is $p$). 
Even if this problem would have been proven in \P, this does not imply that \DEEP\ in \P, because the input of \DEEP\ are integers
and hence it has logarithmic size (\ie a cycle $C_p$ is represented using $\log_2 p$ bits).
\medskip

The proof
of Theorem~\ref{th:deep} is based on
the following characterization
of the solutions of basic equations on the digraphs
of permutations. In the sequel, we denote 
$\factors{p}=\set{p_1,\ldots,p_r}$ the set of
primes appearing in the prime factor decomposition
$p_1^{h_1}\cdot\ldots\cdot p_r^{h_r}$ of the integer $p$.

\begin{theorem}\label{th:condP}
Let $p, q, n$ be positive integers and let $p_1^{h_1}\cdot\ldots\cdot p_r^{h_r}$ (resp., $q_1^{k_1}\cdot\ldots\cdot q_t^{k_t}$) be the prime factor
decomposition of $p$ (resp., $q$).
%the prime decompositions of $p$ and $q$ be as follows:
%\begin{align*}
%p&=p_1^{h_1}\cdot\ldots\cdot p_r^{h_r}\text{ with }\factors{p}=\set{p_1,\ldots,p_r}\\
%q&=q_1^{k_1}\cdot\ldots\cdot q_s^{k_s}\text{ with }\factors{q}=\set{q_1,\ldots,q_s}%\\
%n&=n_1^{m_1}\cdot\ldots\cdot n_t^{m_t}\text{ with }\factors{n}=\set{n_1,\ldots,n_t}\enspace.
%\end{align*}
The equation 
$
C_p \cdot \sum_{i=1}^l C_{z_i}=nC_q
$
has solutions (the values $z_i$ are the unknowns and need not to be distinct) if and only if the two following conditions hold:
\begin{enumerate}
\item $p$ divides $q$;
\item $e$ divides $n$,
% for all $i\in\set{1,\ldots,r}$
%either $p_i^{h_i}\in\set{q_1^{k_1},\ldots, q_s^{k_s}}$ or $p_i^{h_i}$ divides $n$.
\end{enumerate}
where $e=\prod_{v\in E} v$ with $E=\set{q_j^{k_j}\;|\;\exists i\text{\ s.t.\@\ } p_i=q_j\text{ and }h_i<k_j}$.
\end{theorem}
\begin{proof}
The left part of equation can be rewritten as  $\sum_{i=1}^l C_p\cdot C_{z_i}$. Hence, by
Proposition~\ref{prop:prod}, $p$ must divide $q$. This means that $\factors{p}\subseteq\factors{q}$ and that for any $p_i^{h_i}$ there
exists $j\in\set{1,\ldots,t}$ such
that $q_j\in\factors{q}$ and
$h_i\leq k_j$.

\noindent
($\Leftarrow$) Given the factorization of $q$ and Proposition~\ref{prop:prod}, we can rewrite $nC_{q}$ as follows.
\[
nC_q = C_q \cdot nC_1=C_{{q_1^{k_1}} \cdot\ldots\cdot{q_s^{k_t}}}\,\cdot nC_1=C_{q_1^{k_1}}\cdot\ldots\cdot C_{q_t^{k_t}} 
\cdot nC_1
\]

Suppose now that, for some $i\in\set{1,\ldots,r}$, ${p_i^{h_i}}$ is not contained in $\set{q_1^{k_1}, \ldots, q_t^{k_t}}$. Then, there exists $j\in\set{1,\ldots,s}$ such that $p_i=q_j$ and $h_i< k_j$. Now, if $p_i^{h_i}$ divides $n$, we have
\[
C_{q_1^{k_1}}\cdot\ldots\cdot C_{q_s^{k_t}} \cdot nC_1 = C_{q_1^{k_1}}\cdot\ldots\cdot C_{q_s^{k_s}}\cdot C_{{\basep_i^{\espp_i}}} \cdot \frac{n}{\left(p_i^{h_i}\right)}C_1.
\] 
Repeating the same operation for all $p_i^{h_i}$ not included in $\set{q_1^{k_1}, \ldots, q_t^{k_t}}$, we find a solution since $\sum_{i=1}^l C_{z_i}$ will contain a cycle $C_{z_i}$ (with $z_i$ equal to $q_j^{k_j}$) for any $p_i^{h_i}$ not included, and the set of self-loops.

\noindent($\Rightarrow$)
Let us suppose that the equation $C_p \cdot \sum_{i=1}^l C_{z_i}=nC_q$ holds for some $z_1 , z_2 , \ldots , z_l$. Let us assume that there exists $i$ such that $p_i^{h_i}\notin \set{q_1^{k_1}, \ldots, q_t^{k_t}}$. Then, $p_i^{h_i}$ will be equal to a certain $q_j^y$ with $y < k_j$. This implies that all $z_1 , z_2 , \ldots , z_l$ must contain exactly $q_j^{k_j}$ as a factor, since the $\lcm$ with $p$ must be $q$. Thus we can write:

\begin{align*} 
\sum_{i=1}^l C_{z_i} &= C_{q_j^{k_j}} \cdot \sum_{i=1}^l C_{\frac{z_i}{\left(q_j^{k_j}\right)}}
\\
\intertext{which leads us to the following}
\\
C_p \cdot \sum_{i=1}^l C_{z_i} &= C_p \cdot  C_{q_j^{k_j}} \cdot \sum_{i=1}^l C_{\frac{z_i}{\left(q_j^{k_j}\right)}} 
\\
&= C_{\frac{p}{\left(q_j^y\right)}} \cdot C_{q_j^y} \cdot  C_{q_j^{k_j}} \cdot \sum_{i=1}^l C_{\frac{z_i}{\left(q_j^{k_j}\right)}}
\\
&=C_{\frac{p}{\left(q_j^y\right)}} \cdot q_j^yC_{q_j^{k_j}} \cdot \sum_{i=1}^l C_{\frac{z_i}{\left(q_j^{k_j}\right)}}
\end{align*}
Then, according to Proposition~\ref{prop:prod}, $p_i^{h_i}$ (\ie $q_j^y$) must divide $n$.
\end{proof}

Theorem~\ref{th:condP} provides a condition to decide whether a solution to a basic equation can exist. However, we are not
going to exploit it directly, since it requires the
prime factorization of the integers $p$ and $q$ which
would imply unnecessarily high complexity. In fact, the algorithms we are going to conceive will essentially
use simple arithmetical operations or basic functions on the integers. In this regard,
we consider that the worst-case time complexity of the addition (resp., multiplication) between two
integers is $O(n)$ (resp., $O(n^2)$), where $n$ is the input size in bits. The division between two integers is assumed to have the same complexity
as the multiplication.
Concerning $\gcd$, we consider the classical Euclidean GCD algorithm which has worst-case time complexity $O(n^2)$. In the conclusions, we succinctly discuss how the choices in the implementations impacts the final complexity of our algorithm.
\medskip

To determine the complexity of \DEEP, we need two important lemmas. We emphasize that they will be applied to the peculiar case that we are studying but they are valid for any pair of positive integers.

\begin{lemma}\label{lem:PIF}
Let $p,q$ be two positive integers.
Let
$p_1^{h_1}\cdot\ldots\cdot p_r^{h_r}$ and
$q_1^{k_1}\cdot\ldots\cdot q_t^{k_t}$ be the
prime factor decompositions of $p$ and $q$,
respectively. 
%Let us also define
%$\factors{p}=\set{p_1, \ldots, p_r}$ and
%$\factors{q}=\set{q_1, \ldots, q_s}$.
Finally, let $\mathbf{F}$ be the set of all $q_j^{k_j}$ such that $q_j\in\factors{q}\setminus\factors{p}$. Algorithm~\ref{algobasi} computes $\Pi_\mathbf{F}=\prod_{\mathbf{f}\in\mathbf{F}}\mathbf{f}$ without using the factorizations of $p$ and $q$. Moreover, the worst-case time complexity of Algorithm~\ref{algobasi} is $O(s^3)$, where $s$ is the size of the input in bits.
\end{lemma}
\begin{proof}
Denote $P=\factors{p}$ and $Q=\factors{q}$.
If $g=\gcd(p,q)=1$ (Line $2$), then $P\cap Q=\emptyset$ and  $\Pi_\mathbf{F}=q$. Hence, the algorithm returns $q$
(Line $4$). If $g\ne1$, then taking $q/g$ keeps unchanged
the $q_j^{k_j}$ belonging to $Q\setminus P$ and at the same time decreases of at least $1$ the exponents of
the $q_j$ belonging to $P\cap Q$. Therefore, calling recursively (Line $6$) $\Pi_\mathbf{F}$ with $(q=q/g,g)$
keeps decreasing the exponents of $q_j$ belonging to $P\cap Q$ until they become $0$. At that point, $\gcd(q,g)=1$ and the algorithm exits returning $\Pi_\mathbf{F}$. Remark that at each call of $\Pi_F$, we compute a $\gcd$ (Line $2$) plus a division (Line $6$) which cost $O(s^2)$ in total.
Since the depth of
the recursion is $O(s)$,
the worst-case time complexity for the algorithm is $O(s^3)$.
\end{proof}

\begin{algorithm}
\caption{}%[Compute ${\Pi}_\mathbf{F}$]{$\Pi_\mathbf{F}$}
\label{algobasi}
\SetKwInOut{Input}{Input}\SetKwInOut{Output}{Output}
\SetKwFunction{ComputePF}{Compute-$\mathbf{\Pi_F}$}
\SetKwProg{Fn}{Function}{}{}
\Fn{\ComputePF{p, q}}{
\Input{$p$ and $q$ positive integers}
\Output{$\Pi_\mathbf{F}$}%\Output{$\prod_{\mathbf{f}\in\mathbf{F}} \mathbf{f}$ integer}
\vspace{2mm}
$g \newv{\gets} \gcd(p,q)$\;
\eIf{$g$ == $1$}{ 
		\Return{q}\;
}{
	\Return{${\Pi}_\mathbf{F}(g,\frac{q}{g})$}\;
}
}
\end{algorithm}

\begin{lemma}\label{lem:PIE}
Let $p$ and $q$ be two positive integers such that $p|q$.
Let
$p_1^{h_1}\cdot\ldots\cdot p_r^{h_r}$ and
$q_1^{k_1}\cdot\ldots\cdot q_t^{k_t}$ be the
prime factor decompositions of $p$ and $q$,
respectively. 
%Let us also define
%$\factors{p}=\set{p_1, \ldots, p_r}$ and
%$\factors{q}=\set{q_1, \ldots, q_s}$.
Finally,
let $\mathbf{E}$ be the set of all $p_i=q_j\in\factors{p}\cap\factors{q}$ such that $h_i < k_j$. Then, Algorithm~\ref{algoesp} computes ${\Pi}_\mathbf{E}=\prod_{\mathbf{e}\in\mathbf{E}} \mathbf{e} $ without using the factorizations of $p$ and $q$. The worst-case time complexity of
Algorithm~\ref{algoesp} is $O(s^2\log s)$, where $s$ is the size of the input in bits.
\end{lemma}
\begin{proof}
By executing Lines $2$ and $3$ of Algorithm~\ref{algoesp},
we save in $g$ the product of the factors $q^{k_j}_j$ such that
$q_j\in\factors{p}\cap\factors{q}$. However, the exponents of the factors $q_j$ in $g$ are those of $q$ while to compute
$\mathbf{\Pi_E}$ we need those of $p$. This is the purpose of
the while loop (Lines $5$-$9$). Indeed, by recursively squaring
$d$, it will make the exponents of the factors of $q$ selected in $g$ grow bigger than those in $p$ and hence the $\gcd$ of line $8$ will select the factors and the corresponding exponents
that we were looking for. Concerning the complexity,
it is enough to remark that initialisation part of the algorithm (Lines $2$ to $4$) costs
$O(s^2)$ and that
the number of iterations of the while loop
is $O(\log s)$ in the worst-case with a cost of $O(s^2)$ per each iteration. 
%\sr{Pensi che si capisce facilmente il log log?}
\end{proof}

\begin{algorithm}
%\caption[${\Pi}_\mathbf{E}$]{${\Pi}_\mathbf{E}$}
\caption{}
\label{algoesp}
\SetKwInOut{Input}{Input}\SetKwInOut{Output}{Output}
\SetKwFunction{ComputePE}{Compute-$\mathbf{\Pi_E}$}
\SetKwProg{Fn}{Function}{}{}
\Fn{\ComputePE{p, q}}{
\Input{$p$ and $q$ integers s.\@t.\@ $p|q$}
\Output{$\prod_{\mathbf{e}\in\mathbf{E}} \mathbf{e}$}
\vspace{2mm}
%$i \newv{\gets} 1$\;
$d \newv{\gets} \frac{q}{p}$\;
$g \newv{\gets} \gcd(d,p)$\;
$g' \newv{\gets} \gcd(d*d,p)$\;
\While{$g \neq g'$}{ 
	$g \newv{\gets} g'$\;
	$d \newv{\gets} d*d$\;	
	$g' \newv{\gets} \gcd(d,p)$\;
}
\Return{$g$}\;
}
\end{algorithm}

\begin{theorem}\label{prop:dec-sobfid-P}\label{th:deep}
For any positive integers $p,q$ and $n$, \DEEP\ has worst-case time complexity
$O(s^3)$, where $s$ is the size of the input in bits.
\end{theorem}
\begin{proof}
First of all, let us prove that 
Algorithm~\ref{algosobfidP} solves
\DEEP. Indeed, choose three positive integers $p, q$ and $n$ and assume that
$p_1^{h_1}\cdot\ldots\cdot p_r^{h_r}$ and
$q_1^{k_1}\cdot\ldots\cdot q_t^{k_t}$ are the prime number factorizations of $p$ and $q$, respectively.
Line $2$ checks if $p$ divides $q$ 
which is the first condition of Theorem~\ref{th:condP}.
Let us focus on Line $5$. Computing $\Pi_\mathbf{F}(p,q)$, we get the product of all those factors $q_j^{k_j}$ such that $q_j\in\factors{q}\setminus\factors{p}$.
Hence, dividing $q$ by $\Pi_\mathbf{F}(p,q)$ we get the product of all the prime factors of $q$ such that $q_j\in\factors{p}\cap\factors{q}$.
However, in order to check the second condition of Theorem~\ref{th:condP} we need this quantity but the exponents of $q_j$ must be the corresponding ones in the prime factorization of $p$. This is
computed by calling $\Pi_\mathbf{E}$ at
Line $5$. Then, Line $6$ checks if the product of the factors $p_i^{h_i}\in\factors{p}\cap\factors{q}$ 
for which $q_j^{k_j}$ is such that $h_i<k_j$ divides $n$. We stress that
if the product divides $n$, then each single factor of the product divides $n$.
Hence, Algorithm~\ref{algosobfidP}
answers `yes' if and only if the conditions of Theorem~\ref{th:condP} are verified.

Concerning the complexity, we have that
the divisibility test in Line $2$ can be computed
in $O(s^2)$. By Lemmata~\ref{lem:PIF} and \ref{lem:PIE}, we know that computing
$e$ takes $O(\max(s^3,s^2\log s, s^2))$
that is $O(s^3)$. Remark that the size in bits of
the quantities involved in the calculation of
$e$ is bounded by $s$. Indeed, both $\Pi_F$ and $\Pi_E$ return a divisor of $q$.

Finally, the checks in Lines $2$ and $6$ have complexity $O(s^2)$
since $e$ is a divisor of $p$ and, in its turn, $p$ is a divisor of $q$. We conclude that the time complexity is $O(s^3)$.
\end{proof}

\begin{algorithm}
\caption[]{}
\label{algosobfidP}
\SetKwInOut{Input}{Input}
\SetKwInOut{Output}{Output}
\SetKwFunction{DEEP}{DEEP}
\SetKwProg{Fn}{Function}{}{}
\Fn{\DEEP{p, q, n}}{
\Input{$p, q$ and $n$ positive integers}
\Output{$true$ if $C_p \cdot X=nC_q$ has solutions, $false$ otherwise}
\vspace{2mm}
%\eIf( \hspace{2mm}// $p$ does not divide $q$){$\gcd(p,q)\ne p$}{  
\eIf{$p$ does not divide $q$}{  
	\Return{$false$}\;
}(\hspace{4mm}// $p$ divides $q$){
    $e \newv{\gets} \Pi_\mathbf{E}(p,\frac{q}{{\Pi}_\mathbf{F}(p,q)})$\;
	\eIf{$e$ divides $n$}{ 
		\Return{$true$}\; 
	}{
		\Return{$false$}\;
	}
}
}
\end{algorithm}

We show how the previous algorithms interact to produce
the expected results by the following numerical
example.

\begin{example}%[Algorithm \ref{algosobfidP} to decide if $C^1_{8400} \odot X=C^{6000}_{8316000}$ admits a solution]
Consider the following equation
\[
C_{8400} \cdot X=6000C_{8316000}
\]
where $p=8400$, $q=8316000$ and $n=6000$. 
Let us first see how by means of the presented algorithms that we can verify the conditions of Theorem~\ref{th:condP} without knowing the factorizations of the numbers involved. Later, we will see how the method acts at the level of the factorizations.
\medskip

Since $p$ divides $q$, we want to calculate ${\Pi}_\mathbf{E}(\frac{8316000}{{\Pi}_\mathbf{F}(8316000,8400)},8400)$. Let us begin by considering ${\Pi}_\mathbf{F}(8316000$ $,8400)$. Since $8316000$ and $8400$ are not coprime, the method is iterated, \ie $${\Pi}_\mathbf{F}(\frac{8316000}{\gcd(8316000,8400)},\gcd(8316000,8400))={\Pi}_\mathbf{F}(990,8400).$$ Again, $990$ and $8400$ are not coprime, we call recursively the function ${\Pi}_\mathbf{F}(\frac{990}{\gcd(990,8400)},\gcd(990,8400))={\Pi}_\mathbf{F}(33,30)$ which brings us to ${\Pi}_\mathbf{F}(\frac{33}{\gcd(33,30)},\gcd(33,30))={\Pi}_\mathbf{F}(11,3)$. Since $11$ and $3$ are coprime, the method returns $11$.

Let us therefore study ${\Pi}_\mathbf{E}(\frac{8316000}{11},8400)={\Pi}_\mathbf{E}(756000,8400)$. With $i=1$, $d=\frac{756000}{8400}=90$, $g=\gcd(90,8400)$ $=30$ and $g'=\gcd(90^2,8400)=\gcd(8100,8400)=300$. Since $g\neq g'$, $i$ becomes $2$, $g$ takes value $300$ and $g'$ becomes $\gcd(90^3,8400)=\gcd(729000,8400)=600$. Hence, since $g$ and $g'$ are still not equal, the method continues as follows.
\[ i=3, g=600, g'=\gcd(90^4,8400)=\gcd(65610000,8400)=1200\]
\[ i=4, g=1200, g'=\gcd(90^5,8400)=\gcd(5904900000,8400)=1200\]
At this point,
since $g=g'$, the method returns $1200$. Finally, since $1200$ divides $n=6000$ we know that the equation admits a solution.
\medskip

Let us now see what happens, from the point of view of the factorizations, by applying the previous method. Considering the values in input of this example, the factorizations are:
\[p=2^4 \cdot 3 \cdot 5^2 \cdot 7, \;\; q=2^5 \cdot 3^3 \cdot 5^3 \cdot 7 \cdot 11, \;\; n= 2^4 \cdot 3 \cdot 5^3 \]
so we have $\factors{p}=\set{2,3,5,7}$ and $\factors{q}=\set{2,3,5,7,11}$.
The goal of ${\Pi}_\mathbf{F}$ is to compute the product of all $q_j^{k_j}$ such that $q_j\in\factors{q}\setminus\factors{p}$,
and, in fact, the method calculates the following.
\begin{align*}{\Pi}_\mathbf{F}&(2^5 \cdot 3^3 \cdot 5^3 \cdot 7 \cdot 11,2^4 \cdot 3 \cdot 5^2 \cdot 7)=\\
&={\Pi}_\mathbf{F}(\frac{2^5 \cdot 3^3 \cdot 5^3 \cdot 7 \cdot 11}{\gcd(2^5 \cdot 3^3 \cdot 5^3 \cdot 7 \cdot 11,2^4 \cdot 3 \cdot 5^2 \cdot 7)},\gcd(2^5 \cdot 3^3 \cdot 5^3 \cdot 7 \cdot 11,2^4 \cdot 3 \cdot 5^2 \cdot 7))\\
&={\Pi}_\mathbf{F}(2 \cdot 3^2 \cdot 5 \cdot 11,2^4 \cdot 3 \cdot 5^2 \cdot 7)\\
&\text{\textcolor{white}{white}}\\
{\Pi}_\mathbf{F}&(2 \cdot 3^2 \cdot 5 \cdot 11,2^4 \cdot 3 \cdot 5^2 \cdot 7)=\\
&={\Pi}_\mathbf{F}(\frac{2 \cdot 3^2 \cdot 5 \cdot 11}{\gcd(2 \cdot 3^2 \cdot 5 \cdot 11,2^4 \cdot 3 \cdot 5^2 \cdot 7)},\gcd(2 \cdot 3^2 \cdot 5 \cdot 11,2^4 \cdot 3 \cdot 5^2 \cdot 7))\\
&={\Pi}_\mathbf{F}(3 \cdot 11,2 \cdot 3 \cdot 5)\\
&\text{\textcolor{white}{white}}
\end{align*}
\begin{align*}
{\Pi}_\mathbf{F}&(3 \cdot 11,2 \cdot 3 \cdot 5)=\\
&={\Pi}_\mathbf{F}(\frac{3 \cdot 11}{\gcd(3 \cdot 11,2 \cdot 3 \cdot 5)},\gcd(3 \cdot 11,2 \cdot 3 \cdot 5))\\
&={\Pi}_\mathbf{F}(11,3)=11.
\end{align*}
Once ${\Pi}_\mathbf{F}$ has been calculated, through $\frac{q}{{\Pi}_\mathbf{F}(q,p)}$ we obtain the product of all $q_j^{k_j}$ such that $q_j\in\factors{q}\cap\factors{p}$, \ie $2^5 \cdot 3^3 \cdot 5^3 \cdot 7$. Then, the goal of ${\Pi}_\mathbf{E}$ is to calculate the product of all $q_j^{k_j}$ such that $q_j=p_i\in\factors{p}\cap\factors{q}$ such that $h_i < k_j$. Note that $90=2 \cdot 3^2 \cdot 5$.
\begin{align*}
i=1,& g=\gcd(2 \cdot 3^2 \cdot 5,2^4 \cdot 3 \cdot 5^2 \cdot 7)=2 \cdot 3 \cdot 5, g'=\gcd(2^2 \cdot 3^4 \cdot 5^2,2^4 \cdot 3 \cdot 5^2 \cdot 7)=2^2\cdot 3 \cdot 5^2,
\\
i=2,& g=2^2\cdot 3 \cdot 5^2, g'=\gcd(2^3 \cdot 3^6 \cdot 5^3,2^4 \cdot 3 \cdot 5^2 \cdot 7)=2^3\cdot 3 \cdot 5^2,
\\
i=3,& g=2^3\cdot 3 \cdot 5^2, g'=\gcd(2^4 \cdot 3^8 \cdot 5^4,2^4 \cdot 3 \cdot 5^2 \cdot 7)=2^4\cdot 3 \cdot 5^2,
\\
i=4,& g=2^4\cdot 3 \cdot 5^2, g'=\gcd(2^5 \cdot 3^{10} \cdot 5^5,2^4 \cdot 3 \cdot 5^2 \cdot 7)=2^4\cdot 3 \cdot 5^2.
\end{align*}
\end{example}
\subsection{Some special cases}
In this section we provide some special cases of equations
with constant right-hand side which are more general than
basic equations and have quadratic complexity but they provide either they
only sufficient criteria for the existence of
of solutions or necessary and sufficient criteria
but have quite limited application scope.
\medskip

The proof of the following proposition is trivial since it boils down to remark the the number of the
vertices of the digraphs involved in the left part
of the equation must equal the number of vertices of digraphs in the right part. However, we shall provide an alternative proof to explicit some relations that are discussed right after the end
of the proof.
\begin{proposition}
\label{prop:sum-C_pi^mi-X=C^n_q}
Given $r+1$ positive integers $p_1, \ldots, p_r, q$ and $r+1$ integers $m_1, \ldots, m_r$ and $n$. If the following equation
\begin{equation}
\label{eq:sum-C-mi-pi-X=C-n-q}
 \sum_{i=1}^r m_iC_{p_i}\cdot X=nC_q
\end{equation}
has solutions then $\sum_{i=1}^rm_ip_i$ must divide $nq$.
\end{proposition}
\begin{proof}
Denote $d(q)$ the set of divisors of $q$. A generic solution  to
Equation~\eqref{eq:sum-C-mi-pi-X=C-n-q} (if it exists) has the form $\hat X=\sum_{i=1}^{|d(q)|}s_iC_{t_i}$,
where $t_i$ is the $i$-th divisor of $q$ (divisors are considered ordered by the standard order on integers) and $s_i$ is an integer (possibly $0$). 
%\sr{possiamo ridurre dicendo che i divisori possibili sono quelli piu piccoli di n... cosa ne pensi?} 
% preferisco essere preciso...
We have
\begin{align}
\label{eq:gran-casino}
nC_q=\sum_{k=1}^r m_kC_{p_k}\cdot
\sum_{i=1}^{|d(q)|}s_iC_{t_i}
=
\sum_{k=1}^r\sum_{i=1}^{|d(q)|}
m_kC_{p_k}\cdot
s_iC_{t_i}=
\sum_{k=1}^r\sum_{i=1}^{|d(q)|}
m_ks_i\gcd(p_k,t_i)C_{\lcm(p_k,t_i)}
\end{align}
Since $\hat X$ is a solution it must be either $q=\lcm(p_k,t_i)$ or $s_i=0$. Hence, from $p_kt_i=\gcd(p_k,t_i)\lcm(p_k,t_i)$
and the previous consideration, we get $\gcd(p_k,t_i)=p_kt_i/q$.
If we replace this last quantity in Equation~\eqref{eq:gran-casino}
we get
\begin{align}
nC_q=\sum_{k=1}^r\sum_{i=1}^{|d(q)|}
\frac{m_ks_ip_kt_i}{q}C_q
\end{align}
which holds iff
\begin{equation*}
\sum_{k=1}^r\sum_{i=1}^{|d(q)|}
m_kp_ks_it_i/q=n
\end{equation*}
that is to say iff
\begin{equation}
\label{eq:gran-casino-quasi-fine}
\left(\sum_{k=1}^rm_kp_k\right)\left(\sum_{i=1}^{|d(q)|}
s_it_i\right)=nq
\end{equation}
We deduce that if Equation~\eqref{eq:gran-casino-quasi-fine} has solutions then $\sum_{k=1}^rm_kp_k$ divides $nq$.
\end{proof}
Let us make some remarks concerning the proof of the
previous result.
Assume that $H=\frac{nq}{\sum_{k=1}^rm_kp_k}$ is an integer. Then, the Diophantine equation
\begin{equation}
\label{eq:diopha-H}
\sum_{i=1}^{|d(q)|}
s_it_i=H
\end{equation}
has solutions iff $\gcd(t_1, \ldots, t_{|d(q)|})$
divides $H$, which is always true if $\sum_{k=1}^rm_kp_k$ divides $n$. With
standard techniques (see for instance~\cite[Theorem 5.1, page 213]{niven1991}), one can recursively find the solutions in linear time  in the size of $q$ in bits. We point out that it might happen that all the solutions are
non admissible \ie every solution contains at least one $s_i<0$ for $i\in\set{1,\ldots,r}$. This issue can also be settled in linear time
in the size of $q$ in bits (see the considerations following Theorem 5.1 in~\cite[pages 213-214]{niven1991}). However, the big problem that we have is that we need to compute the divisors of $q$ and, as far as we know, the best algorithms for computing them have worst-case time complexity which is sub-exponential in the size of $q$ (see~\cite{BLP1993} for details). 

In a similar manner as in Proposition~\ref{prop:sum-C_pi^mi-X=C^n_q}, one can derive another necessary condition for Equation~\eqref{eq:sum-C-mi-pi-X=C-n-q}
as provided in the following proposition.
\begin{proposition}
\label{prop:p=q^i}
Given $r+1$ positive integers $p_1, \ldots, p_r, q$ and $r+1$ integers $m_1, \ldots, m_r$ and $n$. If Equation \eqref{eq:sum-C-mi-pi-X=C-n-q}
has solutions then $\gcd(m_1,\ldots,m_k)$ divides $n$.
\end{proposition}
\begin{proof}
Denote $d(q)$ the set of divisors of $q$.
From Equation~\eqref{eq:gran-casino} in the proof of Proposition~\ref{prop:sum-C_pi^mi-X=C^n_q}, it is easy to check that $\hat X=\sum_{i=1}^{|d(q)|}s_iC_{t_i}$ is a solution if and only if
\begin{equation}
\sum_{k=1}^r\sum_{i=1}^{|d(q)|}
m_ks_i\gcd(p_k,t_i)=n\enspace.
\end{equation}
Now, set $S_k=\sum_{i=1}^{|d(q)|}s_i\gcd(p_k,t_i)$ we have
\begin{equation}
\label{eq:sum-m_kS_k=n}
\sum_{k=1}^r
m_kS_k=n\enspace,
\end{equation}
where the $S_k$ are the unknowns.
This last equation has solutions iff $\gcd(m_1, \ldots, m_r)$ divides $n$.
However, this last condition is only necessary since solutions might be non-admissible (\ie one or more of the $S_k$ might be
negative).
\end{proof}
Again, we would like to stress that solutions to Equation~\eqref{eq:sum-m_kS_k=n} can be found in quadratic time in the size of the input giving raise to $r$ equations
\[
S_k=\sum_{i=1}^{|d(q)|}
s_i\gcd(p_k,t_i)
\]
which can be also solved in quadratic
time provided that the divisors $t_i$ of $q$ are known. Hence, once again
the problem of finding solutions
to Equation~\eqref{eq:sum-C-mi-pi-X=C-n-q} can be solved in polynomial time in the size of
$q$ in bits (of course, always knowing the divisors of $q$). However, for some particular values of $q$, Equation~\eqref{eq:sum-C-mi-pi-X=C-n-q} can be solved in a more effective way. An example is shown by the following proposition.
\begin{proposition}
\label{prop:q^t}
Given $r+1$ positive integers $p_1, \ldots, p_r, q$ and $r+2$ integers $m_1, \ldots, m_r, t$ and $n$. 
Let $R=\set{1, \ldots, r}$.
Assume $q$ is prime and that for every $i\in R$ we have $p_i\ne q^t$. Then, the following
equation
\begin{equation}
%\label{eq:sum-C-mi-pi-X=C-n-q}
 \sum_{i=1}^r m_iC_{p_i}\cdot X=nC_{q^t}
\end{equation}
has solutions iff
$\sum_{i=1}^r m_ip_i$ divides $n$
and for every $i\in R$, $p_i=q^{t_i}$ for some integer $t_i<t$. Moreover, if a solution exists, then it is unique.
\end{proposition}
\begin{proof}\mbox{}\\
\noindent$(\Rightarrow)$ Assume that $\sum_{k=1}^lC_{v_k}^{u_k}$ is a solution for the equation. Then, we have
\[
\sum_{i=1}^r m_iC_{p_i}\cdot \sum_{k=1}^lu_kC_{v_k}=\sum_{i=1}^r\sum_{k=1}^lm_iC_{p_i}\cdot u_kC_{v_k}=\sum_{i=1}^r\sum_{k=1}^lm_iu_k\gcd(p_i,v_k)C_{\lcm(p_i,v_k)}=nC_{q^t}
\]
which implies $\lcm(p_i,v_k)=q^t$. Since $q$ is prime, we deduce that either $p_i=q^t$ or $v_k=q^t$. Using the hypothesis we have that $v_k=q^t$
for all $k\in\set{1,\ldots,l}$. Hence,
the solution can be written as $uC_{q^t}$. Now, for every $i\in R$, we have
$\gcd(p_i,q^t)=p_i$. Hence, $uC_{q^t}$
is a solution iff the following holds
\[
\sum_{i=1}^r m_iC_{p_i}\cdot uC_{q^t}=\sum_{i=1}^r m_iup_iC_{q^t}=\left(\sum_{i=1}^r m_iup_i\right)C_{q^t}=nC_{q^t}
\]
which holds iff $\sum_{i=1}^r m_iup_i=u\cdot\sum_{i=1}^r m_ip_i=n$.\\
\noindent$(\Leftarrow)$ It is not difficult to see that if $\sum_{i=1}^r m_ip_i$ divides $n$, then $\frac{n}{\sum_{i=1}^r m_ip_i}C_{q^t}$ is a solution provided that $p_i$ divides $q^t$ which
implies that $p_i=q^{t_i}$ for some $t_i\leq t$.

Concerning the uniqueness of the solution, it is enough to remark that the question is equivalent to asking for solutions of a Diophantine equation on a single variable.
\end{proof}

Clearly, the complexity of the algorithm verifying the conditions of Propositions~\ref{prop:sum-C_pi^mi-X=C^n_q} and \ref{prop:p=q^i}
is quadratic in the size of the input but they are only necessary conditions. On the other hand,
Proposition~\ref{prop:q^t} is both a necessary and sufficient condition, still having a quadratic complexity in the size of the input, but it has a limited applicability because it requires quite specific relations between the coefficients.

\section{Conclusions}
This paper is concerned with \DEEP\ \ie the problem of deciding if a basic equation on permutations has
a solution or not. We show that \DEEP\ has cubic complexity. However, it is clear that with more care in the choice of
the implementation of some of the components of the decision algorithm, we could obtain a time complexity
located between $O(s^{2+\epsilon}\log^k s)$ with optimised arithmetic operations and $O(s^3)$ with naive implementation of arithmetic operations. Our purpose was essentially to prove that the problem
could be solved in polynomial time with a polynomial of \emph{reasonable} degree, but it is clear that when
the instances have large sizes, like in some practical applications, then the shift towards more complex
implementations of the components should be taken into account.

For some other variants of \DEEP, we were able to prove only some necessary conditions which are testable 
in quadratic time. However, they have some interest of their own as they might also be of help in pruning
the search space for software that aims to solve general polynomial equations on functional digraphs such as the software pipeline  proposed in \cite{DENNUNZIO2023101932}. 

%\sr{solito appunto fatto prima... sono ok nell'idea ma la metterei giu in modo diverso}

Several extensions to our results are possible. The most natural one consists of moving from functional digraphs
of permutations to larger classes of functional digraphs taking into account the new
recent results of \cite{DORE2024114514,DorePPRR24} and \cite{NAQUIN2024114509}.

%A second one would investigate effective algorithms to enumerate solutions of basic equations in an effective way \ie in such a way that producing two successive solutions is polynomial in the size of the first solution. \sr{quindi anche questa frase la girerei diversamente...}

%\sr{trovo che potremmo mettere qualcosa per legare un po di piu queste due parti delle conclusioni... forse un po troppo netto il passaggio?}

Another quite interesting extension would consider general digraphs and not just functional digraphs.
From the work of \cite{CALDERONI2021}, we know that the compositeness testing problem (\ie answering `yes' if the graph in input is the direct product of two other graphs) for general graphs is \GI-hard. Since a basic equation on general graphs can be seen as a variant of the compositeness problem, the computational complexity of solving basic equations for general graphs is expected to be comparable to \GI\ or even harder. 
This motivates the search for
further digraphs/graphs classes for which there exist efficient algorithms for solving basic equations. One starting point could be the class of digraphs with out-degree $2$.

%\newpage
%\begin{equation*}
 %   \sum_{i=1}^S \structure{\chi_i,f_i} \cdot y_i^{w_i}=\structure{\chi_C,f_C} 
%\end{equation*}
%\begin{equation*}
 %   \bigoplus_{i=1}^S \bigoplus_{j=1}^{K_i} C_{n_{ij}}^{p_{ij}} \odot \mathring{y}^{w_i}= \bigoplus_{j=1}^{K_C} C_{n_{Cj}}^{p_{Cj}}
%\end{equation*}
%\begin{equation*}
 %   \sum_{i=1}^S |\chi_i| \cdot  \overline{y}_i^{w_i}=|\chi_C| 
%\end{equation*}
%
\section*{Acknowledgements}
This work was partially supported by:
\begin{itemize}
\item 
the HORIZON-MSCA-2022-SE-01 project 101131549 ``Application-driven Challenges for Automata Networks and Complex Systems (ACANCOS)'';
\item the PRIN 2022 PNRR project ``Cellular Automata Synthesis for Cryptography Applications (CASCA)'' (P2022 MPFRT) funded by the European Union – Next Generation EU;
\item ANR ``Ordinal Time Computations'' (OTC), ANR-24-CE48-0335-01.
\end{itemize}
\bibliographystyle{abbrvnat}

%\bibliography{biblio.bib}

\end{document}